\documentclass[sigchi]{acmart}

\usepackage{booktabs} 

\setcopyright{none}

\begin{document}
\title{Exploring Factors that Influence Connected Drivers to (Not) Use or Follow Recommended Optimal Routes}

\author{Briane Paul V. Samson}
\orcid{0002-0253-452X}
\affiliation{%
  \institution{Future University Hakodate}
  \city{Hakodate}
  \country{Japan}
}
\affiliation{
  \institution{De La Salle University}
  \city{Manila}
  \country{Philippines}
}
\email{b-samson@sumilab.org}

\author{Yasuyuki Sumi}
\affiliation{%
  \institution{Future University Hakodate}
  \city{Hakodate}
  \country{Japan}
}
\email{sumi@acm.org}




\begin{abstract}
Navigation applications are becoming ubiquitous in our daily navigation experiences. With the intention to circumnavigate congested roads, their route guidance always follows the basic assumption that drivers always want the fastest route. However, it is unclear how their recommendations are followed and what factors affect their adoption. We present the results of a semi-structured qualitative study with 17 drivers, mostly from the Philippines and Japan. We recorded their daily commutes and occasional trips, and inquired into their navigation practices, route choices and on-the-fly decision-making. We found that while drivers choose a recommended route in urgent situations, many still preferred to follow familiar routes. Drivers deviated because of a recommendation's use of unfamiliar roads, lack of local context, perceived driving unsuitability, and inconsistencies with realized navigation experiences. Our findings and implications emphasize their personalization needs, and how the right amount of algorithmic sophistication can encourage behavioral adaptation. 
\end{abstract}

%
%

\ccsdesc[500]{Human-centered computing~Human computer interaction (HCI)}
\ccsdesc[300]{Human-centered computing~Empirical studies in HCI}

\keywords{Driving; Navigation Applications; Advanced Driver-Assistance System; Recommender Systems; Navigation Behavior}

\maketitle

\section{Introduction} 
Advanced driver-assistance systems (ADAS) have become ubiquitous in modern vehicles because of the recent developments in communication and sensor technologies. They are primarily developed to improve driving performance, and car and road safety by providing automation and adaptive capabilities to vehicle systems. One of the most widely used tool for driver assistance are automotive navigation systems, which were initially designed to provide digital maps, route guidance for the shortest path to a destination, and traffic incident information \cite{Mikami1978CACS-UrbanControl}. As more private vehicles occupied our roads and more cities are being designed to accommodate and regulate their widespread use, modern automotive navigation systems now also provide information on the cheapest and fastest routes, and traffic condition.

Today, more than half of the world's population call cities their home due to urbanization and a rising middle class \cite{UnitedNations2017}. As we see a consequential increase in car ownership, our efforts in promoting and ensuring sustainable cities are at stake. With dense urban districts and complex road infrastructures, persistent traffic congestion poses a negative effect on our productivity, health, environment, and social equity \cite{Mehndiratta2017}. The worsening traffic conditions have compelled drivers to circumnavigate congested roads and several solutions have been introduced to address this growing problem. Intuitively, cities invest heavily on improving and increasing road network capacity; but adding more links between origin-destination pairs was proven to be counterintuitive and may cause longer travel times \cite{Braess2005,Afimeimounga2005}. 

Another approach was to efficiently manage traffic flow on existing road infrastructures by connecting current fleets to Intelligent Transportation Systems (ITS). Cities have already invested heavily on ITS infrastructure such as toll gantries, adaptive traffic signals, variable-message signs, and traffic detection systems, among others -- all aimed to regulate road use, to capture and provide situational information to drivers, and to redirect them from congested routes. At the same time, in-car navigation and other advanced driver-assistance systems are continually becoming more context-aware -- communicating with other vehicles, the ITS infrastructure, and other smart devices, as well as detecting its immediate environment \cite{Alghamdi2012, Monreal2014, Ali2018}. However in some cases,  in-car navigation systems are barely used and noticed \cite{J.D.Power2012VehicleDeclines}, are becoming too complex to operate \cite{J.D.Power2017ImprovementsFinds}, are not always updated with the latest maps, and sometimes without access to real-time traffic information, which directly impacts their adoption and forcing drivers to find other options.

In the absence and or shortcomings of in-car navigation systems on some vehicle models, smartphone navigation applications such as Waze and Google Maps, have become a preferred alternative for drivers who experience traffic congestion on a daily basis. In the App Annie Rankings \cite{2018GoogleAnnieb}, Google Maps has consistently been the top choice since its introduction of GPS turn-by-turn navigation in 2008. Meanwhile, Waze reported in 2016 that they are already being used in 185 countries by more than 65 million monthly active users \cite{Waze2016DriverIndex}. Other popular navigation applications include HERE WeGo, MapQuest and Bing Maps, but in other countries like Japan, Navitime has been a long time favorite. These navigation applications are free to use and has the latest maps. With the improved sensors in smartphones, these navigation applications started using floating car data from online users to estimate traffic conditions and uses that to suggest optimal driving routes. Maximizing connected drivers, Waze crowd-sources traffic and accident reports, and advisories of police presence, speed traps, and road closures to supplement its turn-by-turn navigation \cite{Levine2014SystemExchange, Valdes-Dapena2016MostDirections}, setting it apart from traditional navigation systems while supporting the notion of navigation as a social activity among drivers and navigators \cite{Forlizzi2010WhereTurn}. At its core, modern in-car navigation systems and navigation applications are routing services, but they are also considered recommender systems because of their sophisticated recommendation engines that use actual and or average road speeds for calculating fastest routes, and learn new routes to suggest to other drivers \cite{2018RoutingServer}. These information on existing road infrastructure and driving behavior have inspired governments to consider their use in influencing future mobility patterns \cite{Ben-Elia2015ResponseReview,Attard2016TheSystems}. 

Our work makes an inquiry into the practices of \emph{connected drivers} -- those who augment their driving with in-car navigation systems and or mobile applications, and are always connected to the Internet. We also sought to understand the human factors behind their (non-)use of and (non-)compliance with recommended optimal routes. In a semi-structured qualitative study with 17 \emph{connected drivers}, we recorded their commute and non-commute trips and found that \emph{connected drivers} mostly do not use and follow recommended routes on daily commute trips, but still leave it on for the duration of the trip. In this paper, we:

\begin{enumerate}
\item Illustrate how \emph{connected drivers} integrate navigation systems and applications into their daily commute and non-commute trips;
\item Describe if, when and where deviations from the recommended routes happened, as well as the reasons why certain navigating decisions were made;
\item Discuss design implications for supporting the navigation needs of a \emph{connected driver}; and 
\item Reflect on how we can design better navigation experiences to support behavioral adaptation.
\end{enumerate}

\section{Related Work}

\subsection{Interacting with Recommender Systems}
With the incredible amount of data from digital and social media, and those from connected devices and sensors in the Internet of Things, recommender systems have been a boon to digital natives in making sense of and discovering new information. This popularity has gained significant attention to its evaluation in HCI, especially for a more user-centric approach. Knijnenberg et. al. \cite{Knijnenburg2012ExplainingSystems} evaluated collaborative filtering recommender systems and found that increased usage is strongly correlated to a positive personalized experience, but their perceptions, experiences and behaviors change over time. These are also influenced by personal and situational characteristics such as age, gender and domain knowledge. Additionally, they found that when users perceive a recommendation set as more diverse, they see it as more accurate and less difficult to choose from. This is echoed by Ekstrand et. al. \cite{Ekstrand2014UserAlgorithms} when they found users choose a system with more diverse recommendations. They also emphasized the importance of building trust in the early use of recommender systems as their results show negative effects of novelty. Comparing between collaborative, content-based and hybrid recommender systems, Wu \cite{Wu2015HybridSystems} found that users mostly preferred recommendation sets that use hybrid filtering. In particular, users see more benefit in recommendations that match their own behavior history (content-based) than those that match the history of similar users (collaborative). Moving to a different type of system, Rong and Pu \cite{Hu2010ASystems} developed a personality-based recommender system and found that novice users had an easier time building their profiles using personality quizzes because it doesn't need much domain knowledge. When users were asked to build profiles for themselves and their friends, they perceived the recommendation for their friends as more accurate. Much of these works have focused on user perceptions and behaviors towards the main approaches to recommender systems with a single criterion for matching, and they have demonstrated user-centric evaluations besides algorithmic accuracy. However, further analysis is needed for the growing number of mobile and ubiquitous recommender systems that incorporate multi-criteria preferences, probabilistic models, and temporal, spatial and crowd-sourced information.

With a focus on GPS devices, Dingus et. al. \cite{Dingus1997a} did camera and instrumented car studies for drivers who use TravTek. They found that older drivers have a difficult time driving and navigating, and despite being more careful, they still made more safety-related errors. Generally, drivers benefited most when using turn-by-turn guidance with voice, resulting to less glances to the device and faster travel times. In their naturalistic field study, most drivers used the GPS device in their rental cars. Al Mahmud et. al. \cite{Mahmud2009UserDrivers} also found old drivers having difficulties with in-car GPS. As a result, they tend to not follow it completely due to reliability concerns and high amount of instructions. On the other hand, the younger drivers were found to be too dependent at times. Lastly, Brown \& Laurier's study \cite{Brown2012TheGPS} documented five problems that drivers usually encounter with their GPS devices and considers it a skilled activity. In order for a driver to have a positive experience and make suitable \emph{instructed actions}, other than giving focus on providing very detailed instructions which can overwhelm and cause more confusion, it is equally important to support the driver's interpretation and analysis of an instruction or new information as they move and figure out what to do next. Clearly, these works have shown how driving and navigating performance is affected by the use of early smartphone, dashboard-mounted and in-car GPS devices. But with a new generation of navigation applications that dynamically adjusts to real-time and historical contextual information, and provides sets of crowd-sourced information, further analysis is needed to see whether there are changes in navigating practice and decision making, and whether they are associated with the type of trip, trip context, and road conditions.

\subsection{Potential for Behavioral Adaptation}
Because of the ubiquity, cost-effectiveness, and positive utility of smartphone navigation applications, there is growing optimism of their potential in improving urban participatory sensing \cite{Silva2013TrafficAlerts,Xie2015AnNetworks,Silva2016UsersOpportunities} and in shaping sustainable mobility patterns among driving citizens \cite{Ben-Elia2015ResponseReview,Attard2016TheSystems}. There are three categories of travel information that can affect travel behavior, namely experiential, descriptive, and prescriptive \cite{Ben-Elia2015ResponseReview}. Experiential information are provided as feedbacks or repeated information from previous experiences, while descriptive information depict current conditions based on historic or real-time data such as estimated times of arrival and traffic conditions. Utilizing experiential and descriptive information, prescriptive information can come as suggestions (e.g. shortest, fastest, and cheapest routes) and or guidance (e.g. turn-by-turn directions). Nowadays, most modern navigation applications provide descriptive and prescriptive information as their main features \cite{Sha2013SocialNavigation}. In Chorus's \cite{Chorus2006TravelReview} and Ben-Elia's \cite{Ben-Elia2015ResponseReview} literature reviews, they have highlighted the extensive focus of recent works on the positive effects of experiential and descriptive information to influence the travel behavior of car drivers. Experiential information has been proven helpful in adapting to uncertain conditions, while descriptive information is particularly valuable in coping with non-correlated and Black Swan events like road accidents and sudden bad weather. However, there is still relatively few studies about the implications of prescriptive information. 

\begin{table*}[t]
	\centering
    \caption{Participant demographic, socioeconomic and driving profiles.}~\label{tab:demographic}
    \begin{tabular}{l l l l l l l}
      & {\textbf{Driving}} & & {\textbf{Monthly}} & & & {\textbf{Driving }}\\
      {\textbf{Participant}}
      & {\textbf{Years}}
      & {\textbf{Occupation}}
      & {\textbf{Income (US\$)}}
      & {\textbf{Nationality}} 
      & {\textbf{Domicile}}
      & {\textbf{Locations}}\\
      \hline
      P1 (F, 20 y.o.) & 1-5 & Student & $<$\$500 & Filipino & Mandaluyong, PHI & Philippines \\
      P2 (M, 20 y.o.) & 1-5 & Student & $<$\$500 & Filipino & Makati, PHI & Philippines \\
      P3 (M, 28 y.o.) & 1-5 & IT Consultant & \$1,000-\$1,500 & Filipino & Taguig, PHI & Philippines \\
      P4 (M, 28 y.o.) & 1-5 & Software Engineer & \$1,000-\$1,500 & Filipino & Makati, PHI & Philippines \\
      P5 (F, 28 y.o.) & 1-5 & Supervisor & $>$\$2,000 & Filipino & Winnipeg, CAN & Canada; USA\\
      P6 (M, 58 y.o.) & $>$10 & Self-Employed & \$500-\$1,000 & Filipino & Makati, PHI & Philippines \\
      P7 (M, 50 y.o.) & $>$10 & Professor & $>$\$2,000 & Japanese & Hakodate, JPN & Japan \\
      P8 (F, 28 y.o.) & 1-5 & Nurse & $<$\$500 & Filipino & Makati, PHI & Philippines \\
      P9 (F, 28 y.o.) & 1-5 & Consultant & $>$\$2,000 & Filipino & Makati, PHI & Philippines \\
      P10 (F, 28 y.o.) & 1-5 & Medical Doctor & $<$\$500 & Filipino & Manila, PHI & Philippines \\
      P11 (M, 30 y.o.) & 5-10 & Sales Director & \$1,500-\$2,000 & Filipino & Quezon City, PHI & Philippines \\
      P12 (M, 20 y.o.) & 1-5 & Student & \$500-\$1,000 & Japanese & Hakodate, JPN & Japan \\
      P13 (M, 20 y.o.) & 1-5 & Student & \$500-\$1,000 & Japanese & Hakodate, JPN & Japan \\
      P14 (F, 42 y.o.) & $>$10 & Pharmacy Assistant & \$1,500-\$2,000 & Filipino & Hakodate, JPN & Japan \\
      P15 (M, 29 y.o.) & 1-5 & Entrepreneur & \$500-\$1,000 & Filipino & Makati, PHI & Philippines \\
      P16 (M, 22 y.o.) & 1-5 & IT Specialist & \$500-\$1,000 & Filipino & Manila, PHI & Philippines \\
      P17 (M, 29 y.o.) & 5-10 & Data Scientist & \$1,500-\$2,000 & Filipino & Caloocan, PHI & Philippines \\
      \hline
    \end{tabular}
\end{table*}

\subsection{Route Choice and Driver's Compliance}
Developers have so far focused on the assumption that drivers would always follow the fastest route to a destination. For most navigation applications, drivers are provided with a number of recommended routes based on a criteria and they can select which one to follow. By default, the fastest route criteria is set unless customizations are made. In the case of Waze, it immediately starts the turn-by-turn navigation and leaves it to the user to check alternative options \cite{Levine2014SystemExchange}. However, this doesn't seem to be the case based on studies examining GPS track data. Zhu and Levinson \cite{Zhu2015DoPrinciple} noticed from GPS tracks that drivers do not always choose the shortest path in their daily commutes. In the follow up work of Tang et. al. \cite{Tang2016AnalyzingData}, some drivers even take a different route each day for their commutes. Recognizing that desired driving experiences have an influence on route choice and vice versa, Pfleging et. al.'s \cite{Pfleging2014ExperienceNavigation} web survey show that the most considered factor for drivers is whether it is the fastest route, but when asked to choose a route from work to home using a prototype navigation screen, 49.1\% chose the fuel-efficient route. Only 18.4\% and 3.5\% chose the fastest and shortest routes, respectively. While these provide rich empirical evidence, it is not clear whether the same prioritization and decision making holds true in real driving scenarios under different circumstances.

Relatedly, Fujino et. al. \cite{Fujino2018DetectingTracks} conducted a more recent study to investigate the phenomena of drivers deviating from the recommended optimal routes of in-car navigation systems and where they usually happen. They analyzed GPS tracks that were collected over 4 years within a 20km\textsuperscript{2} area in Kyoto, Japan. They found that drivers have made significant deviations on intersections with poor on-road signages and those near tourist areas. They also speculated on possible reasons for the deviations based on the physical characteristics of the intersections. While these studies already provide empirical evidence on the surprising route choice and non-compliant behaviors of drivers, none of them had prior knowledge whether the observed drivers used prescriptive information from in-car navigation systems or navigation applications. In the case of \cite{Zhu2015DoPrinciple, Tang2016AnalyzingData, Fujino2018DetectingTracks}, they had no information on the intended route of the drivers nor do they know if the drivers were initially following the guidance of the in-car navigation system used to collect the GPS tracks. Thus, further investigation is warranted to understand why drivers deviated from the recommended optimal routes and whether they chose a recommended route in the first place.

In HCI, Brown \& Laurier's study \cite{Brown2012TheGPS} also noted instances of drivers not following GPS recommendations from their corpus of naturalistic video data. They argue that GPS use is rather a skilled activity as drivers need competency to overcome the \emph{normal, natural troubles} that GPS devices make. Several of these problems such as complex routes, superfluous instructions, map and sensor inaccuracies, and timing of instructions, offer a glimpse as to why GPS recommendations are not followed. Addressing the complex route problem, Patel et. al. \cite{Patel2006PersonalizingRoutes} found that drivers prefer simplified route instructions using familiar landmarks.

As more \emph{connected drivers} use descriptive and prescriptive information from navigation applications and more government stakeholders seek to use them in managing road networks, it is crucial that navigation applications become successful in shaping the travel behavior of connected drivers. Ali et. al. \cite{Ali2018} argues that behavioral adaptation is directly affected by the degree of compliance a driver has with the information provided by navigation applications. Although they are referring to connected vehicle technologies, the same assertion can also be made for navigation applications because they provide the same kind of information. It is worth exploring how we can better utilize descriptive information and present prescriptive information to create navigation experiences that encourages behavioral adaptation.

\section{Method}

\subsection{Participants} 
We recruited 17 driver participants with at least 1 year of driving experience and is using at least 1 navigation application or in-car navigation system through word-of-mouth and social network sharing (See Table \ref{tab:demographic}). We only recruited drivers with at least a year of driving experience as they are likely to be adept in navigation and have acquired preferences (e.g. on safety, road condition, familiarity), but we did not recruit participants that involve driving as their main line of work (i.e. Uber drivers). We also made sure they are not novice users and currently using a navigation application or in-car navigation system as they are likely to have a considerable amount of experience with the features (e.g. turn-by-turn navigation, traffic condition, reporting). We recruited participants from Japan and the Philippines mainly because of their wide exposure to in-car navigation systems (in Japan) and navigation applications (in the Philippines). They also comprise an underrepresented driving population (Filipinos) in literature who may largely benefit from technology improvements. We aim to see common behaviors and factors considered despite the difference in driving culture and technologies used.  

Participants submitted their personal details (i.e. age, sex, occupation, and monthly income range) and driving background using a Google Form survey at the beginning. This allows for an examination of possible motivations for their commute and non-commute trips. We also asked whether they use in-car navigation systems and or navigation applications, and the number of years they have been using them. 

\subsection{Study Protocol}
After answering the pre-collection survey, we conducted a semi-structured qualitative study \cite{Soegaard2013} by recording trips in a naturalistic setup. Extending the scope of naturalistic driving data of Brown and Laurier \cite{Brown2012TheGPS} and Dingus et. al. \cite{Dingus1997a}, we focused on collecting data on the practices of using navigation applications for 3 trip types along with their trip context. We also collected data on whether they chose the recommendation or not, and the factors considered. Recordings were processed and trips were traced to extract instances of deviations. We then did a post-collection interview and used the grounded theory method \cite{Muller2016MachineMethod,Muller2014CuriosityMethod} for the survey answers, video recordings, trip data, in-car conversations and interviews to better understand their navigation practices and reason for route choices, and to uncover their motivations behind deviations made. 

\subsection{Trip Recordings}
Each participant were asked to record at least one instance of the following types of trips: Home-to-Work, Work-to-Home, and Home/Work-to-Unknown. The Home-to-Work and Work-to-Home trips represent their daily commutes. For the Home/Work-to-Unknown trips, the participants recorded their occasional trips to a location they do not usually go to or haven't been to before. 

Inside the participant's vehicles, we attached a commercial dual lens dash camera behind the rear-view mirror to record the changing conditions on the road, and the driver and passenger/s attention. We wanted to capture how a driver and/or a navigator (because it can be someone besides the driver) behaves and what is seen on the road when a deviation happens. The dash camera also recorded the GPS tracks, speed, and in-car conversations. For P1, P2 and P6, a data collector was riding with them to perform shadowing and asked questions as needed. The rest of the participants collected by themselves and were asked to think aloud. Before each trip, participants noted down their origin, destination, reason for the trip or the first activity to be done upon arrival (e.g. attend a meeting, attend family gathering, etc.), and whether it was urgent. We were able to collect 65 trip recordings in total -- 18 work-to-home, 13 home-to-work, and 34 occasional non-commute trips. Among these, 12 trips did not have any deviations, leaving only 53 trips for analysis. 

\subsection{App Recordings}
To keep track of the application behavior and recommended routes, participants recorded the screen of their smartphones with the navigation application open. This allowed us to observe how the driver and/or navigator used the application while navigating. It also allows us to track how the application behaves after every deviation and how the driver adjusts to the changes.

\subsection{Trip Tracing and Processing}
After data collection, we viewed the trip and app recordings and manually traced each trip's actual route taken and the first recommended route using Google MyMaps. We then marked the deviations (if any) made, and the app's recommended rerouting after each deviation. Trip durations and total distances of both actual and recommended routes were computed using the traces on Google MyMaps. We initially wanted to quantify gaze from the in-car videos but almost all drivers were using voice guidance. We did not pursue it but still observed where they paid attention to. 

In preparation for the post-collection interviews, we synchronized the dash camera and app recordings, and made clippings that focused on parts of the trips when deviations happened. We included 10 seconds of video before and after each deviation to provide more context.

\subsection{Post-Collection Interview}
In a separate interview after the data collection and processing, we first asked the participants about their daily routines and their motivations and experiences in using navigation applications. We then presented their trip traces and synchronized clippings when deviations happened. The interviews lasted between 60 to 90 minutes on average, and were focused on recollecting navigation experiences and examining the motivations behind choosing a route, the deviations made(if any), perceptions about the road conditions and recommended routes, as well as other observations and insights from the videos. 

\subsection{Data Analysis}
Finally, we did an iterative coding and thematic analysis of the interview answers, in-car conversations and videos. We did a pilot analysis with 7 participants while the 10 other participants are still collecting. We achieved saturation after only a few new codes and themes were generated for the next 10 participants.

\section{Findings}

\subsection{Navigation Practices}

\begin{figure}[h]
  \centering
  \vspace{-.20cm}
  \centerline{\includegraphics[scale=0.9]{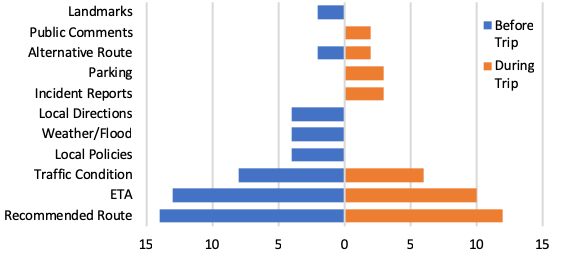}}
  \vspace{-.30cm}
  \caption{The number of participants who accessed certain types of information before and during their trips.}
  \label{fig:info_sought}
  \Description{A horizontal bar graph with the vertical axis in the middle 0 value. The bars going from the middle to the left shows the number of participants who used a particular information before a trip, while those going to the right show numbers for information use during a trip. Most measurements are represented by traffic condition, estimated time of arrival and recommended route.}
\end{figure}

First, we want to investigate the applications used by the drivers, the information they sought, and the order by which the information were used. For this, we looked into the answers from the pre-collection questionnaire and compared it with the recordings and answers to the post-collection interview. We also used trip and app recordings to see associations with the type and purpose of trip. 

\subsubsection{Applications and systems used} 
In daily commute trips, Waze is primarily used when drivers have previous experiences of traffic congestion along their regular and familiar routes (H2W=66.7\%, W2H=69.2\%). They see Waze as an authoritative application especially when they have a clear intention to avoid being late or heavy traffic conditions. Even though Google Maps also provide turn-by-turn navigation and live/historical traffic information, drivers still put a lot of weight on the social aspect of Waze wherein other drivers can manually report traffic conditions, accidents, and road closures. Drivers gain a sense of confirmation as Waze shows manually reported traffic conditions to the ones they derive from the GPS tracks of connected drivers (P3, P4, P8). Since the road incident reports can be quite vague, drivers also acknowledge the usefulness of the public comment feature that allows other drivers who have passed by that area to share details about the incident. P6 shares that once when he was stuck near the tail of a standstill traffic, his passenger checked the public comments feature helped to get real-time updates from the drivers near an accident. It helped him decide whether he should wait longer or start finding other options. 

For short commute trips that doesn't have many alternative routes and doesn't normally experience significant traffic congestion, P5 opt to use Google Maps instead. She expects to see her regular route as the recommended route by the application and just checks the estimated time of arrival. Additionally, she shares that because Google Auto is installed in her vehicle, she prefers to use Google Maps because she can view the route guidance in a wider screen compared to her smartphone. 

Participants from Japan (P7, P12, P13, P14) were primarily using in-car navigation systems because of its ubiquity in most Japanese vehicles. Aside from the provided basic navigation features and digital maps, they are also connected to the local intelligent transportation systems. P13 shared that in one of his previous trips, his in-car navigation system provided a traffic advisory because of an accident in the national highway. It guided him to leave the national highway using the nearest exit.  

In places where the drivers in Japan (P7, P12, P13, P14) drove in, they did not experience any heavy traffic thus, they were not so compelled to download and use another navigation application. However in one of P14's recorded trips, she used and followed Waze when her in-car navigation system started giving incorrect directions. She was noticeably surprised when the in-car navigation system guided her to a direction that's opposite from the destination. She still made the turn as guided by the system but she had already asked one of the passengers to look for the next turn. The passenger then used Waze. P12 particularly used Waze in one of his occasional trips because it shows the location of speed cameras. He found it very useful especially when driving in an unfamiliar location. He shares that this is not provided by his in-car navigation system. 

Other than those mentioned above, drivers also sought information from social networking sites (e.g. Twitter and Facebook) to check traffic and incident updates from their friend networks and the pages of local transportation agencies (P3, P4, P6). They access these sources to augment the information that is not yet provided by in-car navigation systems and navigation applications. 

\subsubsection{Information Sought}
From the interviews and in-car conversations, we looked into the number of times that the participants mentioned each type of information as part of their trip planning and navigation (Figure~\ref{fig:info_sought}). Three participants (age=28-29 y.o.) who have at least 5 years of continued application usage seek at most 7 of these, while the two youngest participants (age=20 y.o.) only check the ETA.

Drivers were mostly checking the estimated time of arrival of the recommended routes, the roads they needed to take, and the traffic condition as their main criteria for choosing a recommended route to follow. Some of the drivers also checked incident reports and updates (P4, P6) to know how much longer they needed to wait in congested roads. 

Drivers were also seeking localized and contextual information such as transport policies (e.g. travel demand management policies, truck ban hours) and flooding (P3, P4, P8). Common to Philippine metropolitan areas, travel demand management policies disallow certain vehicles to use public roads on specific time periods, and it can differ per city. P4 sought this information because he wants to know if he needs to leave earlier than usual to avoid getting apprehended or not use his car at all. Although some participants explicitly shared that they do not actually seek for this information anymore (i.e. P15, P16, P17) because they only memorized it once and doesn't change. However, we see this information useful for transport network vehicle (i.e. Uber, Lyft, Grab) drivers who take passengers to unknown destinations, across cities. In one instance shared by P6 as he was riding an Uber, the driver was apprehensive in crossing another city as recommended by his Waze application because the driver was not sure whether he's allowed or not. That city had a completely different travel demand management scheme as the rest of Metro Manila. Lastly, P7 shared that during winter, he is seeking local information about roads that are not too slippery and safe to drive on, especially because the main roads are where most cars will go. 

For longer and or occasional trips, drivers were also seeking information about familiar landmarks (P3, P4), good parking spaces and local directions. While in-car navigation systems and navigation applications can provide these information, drivers still seek the knowledge of a local person that knows the ins and outs of an unfamiliar place. 

\subsubsection{Usage behavior}
Drivers have been observed to have different behaviors in accessing information and using these to decide which route to take. 

Before starting their daily commute trips, drivers first check the estimated time of arrival (ETA) of the recommended route. They want to have a quick overview of how long it will take them to get to their destinations. Then, they check their familiarity with the roads that were recommended. They usually check how close it is to their regular routes. If it is completely new to the drivers, they check the alternative recommendations and see if their regular route is included. They check the differences between the estimated times of arrival and decide based on a criteria. If they are leaving very late and or in a rush, they only check the ETA (P4, P10). 

During the trip, drivers start the turn-by-turn navigation but only some of them chose to follow it. For example, P10 still follows her regular route to work but still keeps Waze on to get traffic updates. However in the case of P8, she shares that she always follows the suggested route.

When they suddenly experience slowing down due to unexpected traffic build up, they first check what caused it using the navigation application. If there are no reports on the application, they sometimes check Twitter and or Facebook (P3, P4). For alone drivers, they only get to check this information once they are slowing down or in a complete stop (P4, P17). But as passengers and navigators, they tend to check why there's a sudden slow down in traffic and try to look for possible alternative routes (i.e. P3, P4, P6, P16, P17).

For shorter trips to unknown locations, they only used one tool for route guidance. For longer trips, some participants use a mix of applications to plan and navigate. For instance, P3 and P4 shared that they use Google Maps for planning the trip and Waze during the actual trip. Using Google Maps, they looked for landmarks that they can use during the trip and familiarized themselves with the area. And then during the actual trip, they have Waze or Google Maps turned on from the beginning, but leave it idle. They would start to carefully listen to the directions when they already reach a point that they are unfamiliar with (i.e. P4, P15, P17). This supports Patel et. al.'s findings that drivers preferred routes that use familiar landmarks over very detailed turn-by-turn instructions \cite{Patel2006PersonalizingRoutes}. 

In some trips, they switched to another application because of unreliable or missing information. For example in P12's trip, they stopped following the in-car navigation because its map is not updated with the new roads. They then switched to Waze.

\subsection{Route Choice}

\begin{figure}[h]
  \centering
  \vspace{-0.10cm}
  \centerline{\includegraphics[scale=1.1]{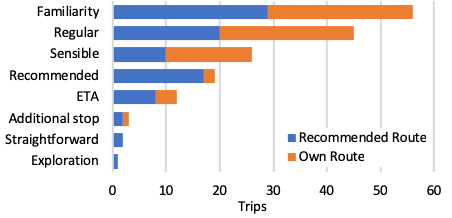}}
  \vspace{-0.30cm}
  \caption{The factors considered for route choice and the number of trips that used them when they chose their own or a recommended route.}
  \label{fig:reason_route_choice}
  \Description{A horizontal stacked bar graph with most values represented by familiarity, regular routes and sensible routes. The number of trips that followed a driver's own route is stacked on top of the trips that followed the recommended route.}
\end{figure}

We also wanted to investigate whether our participants chose to follow the recommended routes given by the applications and in-car navigation systems that they use. We used the app recordings to see how they engaged with an application before a trip. We checked if the destination and agenda upon arrival plays a role. We also analyzed what and how the descriptive and prescriptive information provided were used. 

After analyzing the trip recordings, we found our results consistent with the findings of Zhu and Levinson \cite{Zhu2015DoPrinciple}, and Tang et. al. \cite{Tang2016AnalyzingData}. Our 17 participants chose a route that is not the shortest nor fastest, as computed, in at least one of their recorded trips. At the beginning of each trip, participants decided to use their regular routes in 28 trips (43.1\%), where the occasional non-commute and home-to-work trips each comprised 42.9\%, and 14.3\% were work-to-home. On the other hand, 37 trips (56.9\%) decided to follow recommended routes at the beginning. Majority or 59.5\% of those trips were occasional non-commute, while the work-to-home and home-to-work trips comprised 24.3\% and 16.2\%, respectively. While this contrasts the low preference of drivers for fastest and shortest routes in Pfleging et. al.'s \cite{Pfleging2014ExperienceNavigation} study, this was mainly because Waze and Google Maps do not have options available for eco-routes while the in-car navigation systems used does not make that option apparent to the participants.

Figure~\ref{fig:reason_route_choice} shows how many trips used a which factors to make a route choice decision. In majority or 65\% of the recorded trips, participants considered 3 factors, with familiarity as the most used factor. And while 56.9\% of trips used the recommendation at the beginning, only 21.6\% chose them because of fast ETA. This contrasts the high importance rating of the fastest route factor in the work of Pfleging et. al. \cite{Pfleging2014ExperienceNavigation}.

Before starting their daily commute trips, most participants checked the estimated time of arrival (ETA) and their familiarity with the roads in choosing a route to follow. When they had an important agenda (e.g. meetings, parties) and they were already running late, they chose the fastest recommendation of the application without consideration of familiarity (i.e. P4, P8). For P17, he always turns on the application and follows what recommendation is given. Sometimes, he would inspect the first few roads to decide otherwise.  

When some participants were leaving early and not in a hurry, they always compared the ETA of their regular route with the fastest recommendation. They would chose their regular routes over the fastest recommendation if the time difference is negligible. For instance, P15 shared that he would choose a new recommendation from Google Maps when it is at least 10 minutes faster. But when it is only 2-5 minutes faster, he would still choose a familiar or his regular route. Other participants shared that they would choose a recommended route as long as it has less traffic congestion (i.e. P3, P15), shorter distance (i.e. P3, P5) and straightforward paths (i.e. P8, P14). If some parts of the recommendations do not fit their criteria, they would make a decision to not follow it completely and rely on their own knowledge. 

For occasional non-commute trips, participants chose routes with familiar landmarks (P3, P4), roads familiar to them (P5, P6, P7), and routes suggested by friends living near their destination (P8, P9). For completely new destinations, most participants would follow the application or in-car navigation system completely.

Interestingly, some participants have other reasons for picking a route. For example, P6 shared the he once chose a route with a gas station along the way because they are taking a long trip while P14 chose a route with a specific restaurant along the way because they haven't eaten lunch yet. Other reasons include the need to visit convenience stores (P6, P7) and toilets (P13), and to drop off passengers on the way to work (P6).

Surprisingly, we also found that some participants will open their applications but choose not to follow whatever the application recommends, especially for commute trips. P9 shares that \emph{"In fact, I have self-awareness that in those moments that I know I can, I try to not [follow]."} She doesn't want to be too dependent on the application as she feels that \emph{"whenever there are cases that I cannot use it, I feel incapacitated."} Other participants like P6 shares that most of the time, he just takes his regular route and leave Waze on because he believes that it can learn his regular route. However, even after some months of doing so, the application still doesn't give his regular route as the first recommendation. This non-compliant and non-use behavior supports the findings of Al Mahmud et. al. that some drivers choose not to be too reliant on GPS devices because they know that it can make mistakes and they still have to make their own judgments \cite{Mahmud2009UserDrivers}. 

\subsection{Deviations}

\begin{figure}[h]
  \centering
  \vspace{-0.20cm}
  \centerline{\includegraphics[scale=0.8]{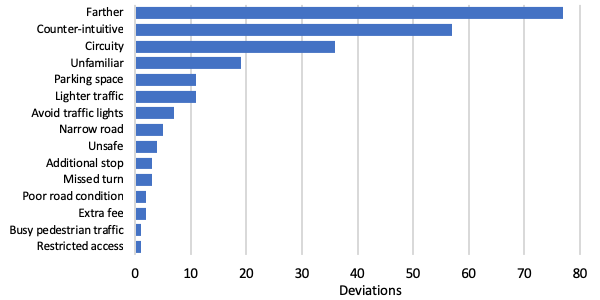}}
  \vspace{-0.30cm}
  \caption{The factors for deviation and the number of deviations they caused.}
  \label{fig:reason_devs}
  \Description{A horizontal bar graph with most values represented by the factors farther, counter-intuitive and circuity.}
\end{figure}

In understanding the motivations behind the deviations, we analyzed the videos, trip data and trip traces to see if any were deliberate or missed turns, and whether they were based on prior knowledge, on information from applications or situational awareness. During the 65 trips, participants deviated 153 times in total. They did it 39 times for home-to-work (M=2.17, SD=5.07), 30 times for work-to-home (M=2.31, SD=6.19), and 84 times for occasional non-commute trips (M=2.47, SD=1.65). 38.5\% of them were single deviations made near the beginning or end of trips, while the extreme cases (3.1\%) made 14-15 deviations. However, there is no clear connection between the types of trips and the number of deviations made. 

Comparing the estimated travel time of the recommended routes and the actual travel times, deviating at least once made the trips longer by an average of 3.11 minutes (N=53, SD=12.35). When only 1 deviation was made, travel times increased by an average of 0.13 minutes (N=25, SD=8.72), and 1.07 minutes (N=45, SD=10.24) for up to 4 deviations. In extreme cases of more than 5 deviations, an average increase of 14.63 minutes (N=8, SD=17.21) was experienced. Although none of the drivers perceived their trips to be longer nor farther after making deviations, this shows that travel time can get worse as more deviations are made.

Looking at trip purpose and urgency, participants made an average of 8 deviations (N=4, SD=5.07) for non-work but urgent trips like catching a flight or appointment, and attending a gathering. When they had to arrive urgently at work, their deviations also increased to an average of 3 deviations (N=12, SD=2.26). But even in non-urgent situations, participants also made more deviations especially when they will only rest (N=13, M=3, SD=3.78) and do leisurely tasks or tours (N=11, M=3, SD=2.5) at the destination. By going through the post-collection interviews, participants revealed various reasons why they deviate from the recommended routes that they choose (Figure~\ref{fig:reason_devs}). In 50.98\% of deviations, more than 1 factor was cited.

\subsubsection{\textbf{Previous Experiences (1.31\%)}}
Participants were mostly deviating from the recommended routes because of their unfamiliarity with some of the roads. This was commonly observed on home-to-work and work-to-home trips where the drivers were recommended fastest routes but were not particularly in a hurry to get to their destinations. For instance, P12 was observed to follow the same path from their hotel to a museum because that was the same path they took when they got to their hotel the previous day. On the other hand, P7 chose to continue on an unfamiliar part of the recommended route because \emph{"This is new to me ... but it seems reasonable because I do not have to make a U-turn."} -- P7

Some participants also deviated because of their past experiences with long waits on traffic lights. In one of his non-commute trips, P4 decided to make an early left turn from the main road, instead of going straight, because \emph{"the next big intersection has a traffic light and I know that's going to take long."} P6 also shares a similar practice when he is recommended to take a main road with traffic lights on its every intersection. \emph{"I just use the smaller road parallel to the main road because it it doesn't have any [traffic light] at all."} -- P6

Participants also consider their negative experiences with past recommendations. P17 shares that he deliberately deviates from a specific road whenever it is recommended. \emph{"Usually, I do not take [street name] ... I opt to go with [another street name] route. I really inspect the route given because I'm avoiding a certain road ... I do not like [taking] [street name] because it's a small road, and when traffic starts, it really regresses along the way. I do not want [to take] it anymore ... It already happened before that Waze asked me to go there and I ended up being stuck there. It happened a lot of times."}  -- P17

\subsubsection{\textbf{Situational Awareness (33.99\%)}}
In most situations, drivers were in situations wherein they have to make quick decisions when their expectations (based on the information that the applications and systems have provided) do not match what they see on the roads. For instance, P6 chose not to follow the next turn recommended by the Waze application because the traffic condition on that road was equally bad as the road he's currently in. Based on perceived road conditions, participants deviated 48.48\% of the time from recommended roads with \emph{medium} traffic conditions to \emph{light} ones and always from recommended roads with \emph{heavy} traffic to \emph{medium} and \emph{light} ones. 

P17 made a similar deviation when he was asked to take a circuitous route through small residential roads, just to return to the road he's currently in. He made a decision to not follow because the traffic is already free flowing on the main road, and not as bad as what is shown on the application. Representing majority of trips with single deviations, participants also deviated near the end of their trips when their initial parking spaces were already full and they had to look for other locations, which was consistent with the findings of Fujino et. al. \cite{Fujino2018DetectingTracks} and the \emph{destination} problem in Brown and Laurier's \cite{Brown2012TheGPS}. This was also the case at the beginning of their trips when they leave their parking spaces.  

Other participants also cited instances when they were directed to gated communities with restricted access, and roads that were unexpectedly blocked. Because their applications were not updated with such information, they just made a conscious decision to take another route and waited for the application to re-route. 

\subsubsection{\textbf{Perceived Driving Suitability (26.80\%)}}
Some participants shared that they did not feel comfortable driving through some of the recommended roads. For instance, P7 was observed to not take a shortcut suggested by the application because \emph{"This is a kind of shortcut but this is a narrow road and it is [a] good route for familiar drivers ... Local familiar drivers and many local drivers tend to use this route but this is narrow and ... it is not so good in dark situation[s]. It is very narrow and [a] very local road ... and usually there's no other walkers [t]here. But if there is, it is very dangerous."} P13 shares the same sentiment when he was asked to take a narrow back street from the hotel in one of their recorded trips. He shared that \emph{"It's a very small road. I do not like to drive on a small road. We're using a rental car, so it's very dangerous."}

Other participants shared instances when they were directed to busy streets and deviated from it. P4 shared one instance when he was recommended to a residential road and he deviated because \emph{"... there's so much pedestrian foot traffic there ... there were also tricycles ... on the same road, it's two-way, so there were also [cars] driving on the opposite direction [but it's narrow] ... so you really have to give way and wait sometimes."} Despite being a more experienced driver, P7 was also observed to deviate from busy main roads especially when going home. \emph{"this route is main road, so then I do not have to ride on that main road just for going to my home."} -- P7

Lastly, mostly female participants (P1, P5, P8, P9, P10, P14) and P7 shared instances of deviating from recommended roads because of poor street lighting conditions, especially in the evening. 

\subsubsection{\textbf{Practicality and Sensibility (94.12\%)}}
Participants were also observed to follow more practical and sensible routes, which goes against most of the recommendations of Waze. Because it was before rush hour and there was still no traffic congestion, P6 was observed to deviate from the recommendation of Waze to take the tolled expressway. Instead, he took a smaller local road, running parallel to the expressway. He argued that since he was not in a hurry, and even if he was, he did not take the tolled expressway because there was no traffic congestion yet. Despite having an option to avoid tolls in the application, he did not enable it at the beginning of the trip because he didn't know the traffic situation until he was near that turn. There was also no way for him to turn it on during the trip as he was already driving. P12, P13 and P14 also showed this behavior when they were recommended to take tolled roads, primarily because of unnecessary extra cost when there is no traffic congestion to beat and they were not in a rush. Aside from P14, the three are students (P12, P13) and self-employed who earn between \$500 to \$1,000.

Other participants deviated because they found some recommendations farther, circuity, winding, and counterintuitive. It's also in these rare cases wherein participants made a trade off to transfer from recommended roads with \emph{light} traffic to \emph{medium} and \emph{heavy} traffic roads (2.65\% of the time), and from recommended \emph{medium} traffic roads to \emph{heavy} traffic ones (6.06\% of the time). For instance, P3 was recommended to take a route that was \emph{"in terms of distance ... when I turn right up to [name of flyover], it is really far."} Instead, he took a route that was comparatively shorter but took longer because of the traffic congestion. Similarly, P4 deviated from a recommendation because it was almost twice as far as his regular route. The application's estimated time of arrival was around 19 minutes and the regular route he took was around 15 minutes only. It seems that the application suggested the longer route because one segment of his regular was showing red in the application, meaning there is reported heavy traffic. However, when he was already at that road segment, he shared that \emph{"surprisingly, well not surprisingly, it was okay ... it's like when I took that road, what I usually take, it was okay. It was free flowing."} Upon analyzing the trip recording, we found that one reason that it was reported as heavy traffic and avoided by the application might be the long wait at the traffic light. But unlike the earlier scenario where P4 avoided the traffic light, this time he didn't because the recommendation was twice as long but only shortens the travel time by around 2 minutes, as indicated by the application. 

Some participants, like P16 and P9, were observed to deviate from roundabout, circuitous recommendations because they saw that they can easily make U-turns. 

\subsubsection{Missed Turns}
While most of the deviations were deliberate, a number of them were actually missed turns due to late, missing, complex and vague instructions. For instance, P15 shared that when he was instructed to turn right in 100 meters, he was not really sure which corner it was because there 4 consecutive corners that were very close to each other. He ended up missing the correct corner to turn to.  Another instance was when P9 was asked to go straight thru an intersection, she couldn't because there were already concrete barriers. She shares \emph{"I was stuck on the left lane and required to turn left because I didn't receive instructions to stay in the middle or right lane ... It was also difficult to cut past the trucks on the middle lane. I stayed on the left lane."}

\section{Discussion}
While it is clear that these applications were mainly designed and developed with the good intention of getting people out of traffic congestion, it is evident from the results that \emph{connected drivers} do not always seek that prescriptive information from navigation applications and in-car navigation systems. For completely unknown destinations, their recommendations made much sense and participants showed high compliance because they do not have prior knowledge to compare with. So they tend to rely on it rather than question its validity. However in most cases during commute trips, they sought traffic and route information relevant to the ones they regularly take. A few of the participants followed whatever is recommended (i.e. P8, P17), many followed recommendations when it matches familiar or regular routes, while some put some constraint on their choices (i.e. P15, P7, P4). These findings cannot be observed in Brown and Laurier's \cite{Brown2012TheGPS} work because they all had their participants follow their GPS device as a condition.

Our list of route choice and deviation factors can be mapped to Pfleging et. al.'s list except for \emph{additional stop}, \emph{parking space}, \emph{restricted access} and \emph{avoiding traffic lights}. Compared to their more generic factors like \emph{least stress}, we expand this work by giving more detailed factors like \emph{circuity} and \emph{counter-intuitive}, which are more useful in coming up with solutions. Surprisingly, their highly rated factor \emph{least fuel consumption} was not considered, along with \emph{no speeding traffic}, \emph{only few trucks}, \emph{low curvature} and \emph{well rated route}, probably because of local considerations. However, this can also be said for factors \emph{avoid traffic lights} and \emph{restricted access} that  only appeared in our findings. However in terms of importance and usage, \emph{familiarity} and \emph{known routes} were mostly considered in 86.15\% (1st) and 69.23\% (2nd) of the trips, whereas in Pfleging et. al., \emph{known route} and \emph{highest driving experience} were ranked low \cite{Pfleging2014ExperienceNavigation}. This shows that even though drivers know there are important factors to consider, their actual use still depends on a trip's purpose, when the choice is being made, and current conditions. 

Drivers also seem to be exhibiting cases of the Einstellung effect \cite{Peterson2018} wherein people are biased towards what they already know, which supports the findings of Patel et. al. \cite{Patel2006PersonalizingRoutes} that drivers prefer personalized routes that include familiar landmarks. We observed this when some drivers made route choices at the beginning of some trips to follow their familiar path even though it was longer and had a later ETA compared to the first recommendation. This was also evident in many deviations wherein they default back to familiar roads when they are about to follow the recommended, yet unfamiliar routes of the application. In the end, they were willing to trade off shorter travel times and distances just so they can be at ease with their navigation choices.

However, if we observe how navigation applications and in-car navigation systems behave, despite considering traffic conditions in their recommendations, they still lack the personalization and sensibility that drivers desire. And quite surprisingly, this caused some participants to completely disregard the recommendation, leave the application on, and go on their own way, hoping that it will learn what it doesn't know yet. But such applications do not learn routes for a single user only. It learns and identifies the best new routes that will be recommended for everyone. This driver behavior and expectation supports Wu's \cite{Wu2015HybridSystems} finding that users have high positive perception when recommendations are matched with their own behavioral history rather than the history similar users. It then raises the question of how much personalization and history is needed.

Finally, it was also observed from the trip recordings that such applications, especially Waze, aggressively recommend and reroute to faster directions for the smallest of gains. And for some participants (i.e. P8, P14), it can be annoying. However, we also found some participants like P17 and P9 who completely understood how such applications work and tend to regard such behavior in a positive way.

\section{Design Implications}

\subsubsection{\textbf{Make Uncertainty Visible}} 
Given the probabilistic and crowd-sourced nature of information shown and used for recommendations on modern navigation applications, there is a tendency for traffic conditions and reports to be unreliable and outdated. This is due to the open problems on data sparsity and in ensuring the integrity of collected reports \cite{Attard2016TheSystems,QingYang2015TowardNetworks,Vyroubal2016MobileSystems}. Because of this concern, we found that drivers were starting to ignore these descriptive information and rely on previous experiences, causing a number of deviations. Although the drivers are unlikely to totally disregard their utility, it is still important to be transparent with the nature of the data we present to users. This can be implemented by considering the uncertain and decaying quality of the crowd-sourced information and try different visualization strategies for improved decision quality. For example, Waze consistently displays a heavily congested road in red and after a few minutes (decay), it either disappears or changes color based on new information. Applying our recommendation, traffic-indicator colors can slowly fade as time passes until an updated information is ready which allows drivers to act properly on information posted minutes ago. For this, we can explore the implementation of value-suppressing uncertainty palettes \cite{Correll2018Value-SuppressingPalettes}, and or Fernandes et. al.'s \cite{Fernandes2018UncertaintyDecision-Making} dotplot or CDF plots which was already tested in a bus transit application. However, as navigation choices are made very quickly, this has to be evaluated for time-critical tasks and prolonged use. Drivers were also found to rely more on voice guidance during trips, so developers may also consider translating these uncertainty information to voice prompts.

\subsubsection{\textbf{Provide Real Personalization}} 
Drivers are idiosyncratic and yet, existing applications still show the fastest route by default. This was evident when only 18.4\% of all trips and 21.6\% of those who followed recommended routes considered a fast ETA for route choice. It is also worth noting that in some of the trips, deviations were clustered on certain areas because their applications assume that the drivers just missed turns and needs to be rerouted back to the recommended route. However, drivers were already deliberately ignoring those, either due to a new route they chose on their own or annoyance \cite{Mahmud2009UserDrivers}. While it is difficult to define a concrete set of conditions that will satisfy their needs, applications can start by learning a driver's mostly used routes and frequently visited landmarks which has been proven to improve user perception \cite{Patel2006PersonalizingRoutes,Wang2014HierarchicalNavigation,Wu2015HybridSystems}. Future navigation applications can show the estimated time of arrival, traffic condition and reports on their mostly used routes so they can properly decide whether they should take a better and new alternative or stick with their regular. Applications may also offer a way to detect when a driver already dislikes the recommended route after a number of deviations, either automatically, by subtle voice commands \cite{Sakamoto2013VoiceManipulation}, quick touch interactions, or a combination of these.

Currently, navigation applications know a lot of about the spatial context of the driver. However, drivers were found to make different route choices, and even make deviations, depending on the type of trip, purpose, and urgency. Some of them also shared their desire to explore scenic routes or routes that will allow them to discover new places or stores along the way \cite{Quercia2014}. Waze and Google Maps already allow integration with personal calendars so that they can make quick searches if the location of the calendar event is already provided. They also allow certain locations to be tagged as \emph{home} and \emph{work}. Future navigation applications may maximize these information and offer drivers to define the intent behind the trip on top of knowing the name of the event. For example, if the driver search directions for tourist destinations, it can infer from the locations that the driver is sightseeing and recommend routes that are scenic and less congested, to maximize the experience. Applications may also use the \emph{home} and \emph{work} tagged locations to offer better recommendations. For example, drivers going home may be recommended straightforward and less stressful routes, which support a common behavior from our findings.

\subsubsection{\textbf{Provide Local Wisdom of Close Network}} 
In uncertain conditions, aside from defaulting to what they are familiar with, drivers are also found to seek information from close friends during trip planning. Some applications already have built-in friend networks while others allow integration with third-party social networks. Hence, applications may offer ways to better maximize these networks to make better recommendations like in the work of Sha et. al. \cite{Sha2013SocialNavigation} where they use \emph{tweets} from nearby vehicles to improve their route recommendations. They may learn the mostly used routes of a driver's close network of friends and prioritize them in the recommendations. One benefit of this is that it provides a sense of community and familiarity. When combined with recommendations based on personal history like in hybrid filtering, user perceptions can also improve \cite{Wu2015HybridSystems}. Additionally, leveraging this information allows the application to improve its recommendations to other drivers who are also going to the same destination. 
 
\subsubsection{\textbf{Be More Persuasive or an Empathetic Other}} 
Our study found that drivers are biased towards what they already know \cite{Patel2006PersonalizingRoutes,Brown2012TheGPS}. This was evident when 86.2\% and 69.2\% of route choices at the beginning of trips mainly considered familiarity and closeness to regular routes, respectively -- a trade off for longer distances and later ETAs. 12.4\% and 37.3\% of deviations where also because of unfamiliarity and counter-intuitiveness. Following the notion of \emph{instructed action} \cite{Brown2012TheGPS}, navigation applications may offer a way to engage drivers in giving route guidance and informing with traffic conditions and crowd-sourced reports, instead of assuming they are docile actors. Antrobus et. al. \cite{Antrobus2017Driver-PassengerSystems} found that collaborative navigation with passengers yield better route knowledge compared to just using SatNav. Thus, applications may offer dialogic route guidance that models collaborative navigation with passengers. Several studies have used a virtual agent \cite{Lin2018Adasa}, an affective robot \cite{Williams2014AffectiveSociability} and even 3 robots in multi-party conversations \cite{Karatas2016NAMIDA:Driver} to reduce cognitive load and distraction. These may be explored so drivers can properly consider options once the rationale behind the recommendations are known. 

\section{Limitations and Future Work}
In this study, participants mostly from the Philippines and Japan, and this bias in the sample may have affected our results. Many of the participants also did not give a complete set of trip recordings for us to analyze. Lastly, we acknowledge that the recorded trips have varying origin-destination pairs thus, controlling some variables like the unknown destination could give us clearer results. In the future, we would like to perform simulation studies to control some variables and gain better insights in explaining the factors that emerged from this work. We also would like to design and test prototypes that try to address the application shortcomings identified in this work to see how they can improve the efficacy and adoption of recommended optimal routes. Future research may also investigate how we can model a driver's intent to deviate from intended routes.

\section{Conclusion}
As governments see potential in navigation applications to shape travel behavior, it is crucial to understand how drivers integrate these in their trips and assess how well the route guidance is complied to and perceived. In this study, we make a first investigation of how users engage with recommender systems enriched with probabilistic and crowd-sourced information. We echo the findings of \cite{Quercia2014, Zhu2015DoPrinciple,Tang2016AnalyzingData,Fujino2018DetectingTracks,Brown2012TheGPS} that drivers do not always choose the fastest route. Further, we uncovered the difference in practice, sets of information sought and used for route choice, and how these are associated with the type of trip, trip context, and driving situations. With all participants making a deviation, we investigated how, when and why they were made. We found that deviations can happen when the recommended route has unfamiliar roads, is impractical and nonsensical, perceived as unsuitable for driving, and the shown descriptive information does not match what they see on the road. Lastly, we present a set of recommendations to design better navigation experiences. These findings and implications emphasize the dynamic and personalization needs of drivers, and provide further evidence that algorithmic sophistication, or less of it, plays an important role in driver compliance and behavioral adaptation. 

\bibliographystyle{ACM-Reference-Format}
\bibliography{Mendeley_Navigo}


\begin{thebibliography}{50}


\ifx \showCODEN    \undefined \def \showCODEN     #1{\unskip}     \fi
\ifx \showDOI      \undefined \def \showDOI       #1{#1}\fi
\ifx \showISBNx    \undefined \def \showISBNx     #1{\unskip}     \fi
\ifx \showISBNxiii \undefined \def \showISBNxiii  #1{\unskip}     \fi
\ifx \showISSN     \undefined \def \showISSN      #1{\unskip}     \fi
\ifx \showLCCN     \undefined \def \showLCCN      #1{\unskip}     \fi
\ifx \shownote     \undefined \def \shownote      #1{#1}          \fi
\ifx \showarticletitle \undefined \def \showarticletitle #1{#1}   \fi
\ifx \showURL      \undefined \def \showURL       {\relax}        \fi
\providecommand\bibfield[2]{#2}
\providecommand\bibinfo[2]{#2}
\providecommand\natexlab[1]{#1}
\providecommand\showeprint[2][]{arXiv:#2}

\bibitem[\protect\citeauthoryear{??}{201}{2018a}]%
        {2018GoogleAnnieb}
 \bibinfo{year}{2018}\natexlab{a}.
\newblock \bibinfo{title}{{Google Maps - GPS Navigation App Ranking and Store
  Data | App Annie}}.
\newblock
\newblock
\urldef\tempurl%
\url{https://www.appannie.com/en/apps/ios/app/google-maps/}
\showURL{%
\tempurl}


\bibitem[\protect\citeauthoryear{??}{201}{2018b}]%
        {2018RoutingServer}
 \bibinfo{year}{2018}\natexlab{b}.
\newblock \bibinfo{title}{{Routing server}}.
\newblock
\newblock
\urldef\tempurl%
\url{https://wazeopedia.waze.com/wiki/Global/Routing_server}
\showURL{%
\tempurl}


\bibitem[\protect\citeauthoryear{Afimeimounga, Solomon, and
  Ziedins}{Afimeimounga et~al\mbox{.}}{2005}]%
        {Afimeimounga2005}
\bibfield{author}{\bibinfo{person}{Heti Afimeimounga}, \bibinfo{person}{Wiremu
  Solomon}, {and} \bibinfo{person}{Ilze Ziedins}.}
  \bibinfo{year}{2005}\natexlab{}.
\newblock \showarticletitle{{The Downs-Thomson Paradox: Existence, Uniqueness
  and Stability of User Equilibria}}.
\newblock \bibinfo{journal}{\emph{Queueing Systems}} \bibinfo{volume}{49},
  \bibinfo{number}{3-4} (\bibinfo{date}{4} \bibinfo{year}{2005}),
  \bibinfo{pages}{321--334}.
\newblock
\showISSN{0257-0130}
\urldef\tempurl%
\url{https://doi.org/10.1007/s11134-005-6970-0}
\showDOI{\tempurl}


\bibitem[\protect\citeauthoryear{Alghamdi and R.Sheltami}{Alghamdi and
  R.Sheltami}{2012}]%
        {Alghamdi2012}
\bibfield{author}{\bibinfo{person}{Wael Alghamdi} {and} \bibinfo{person}{Tarek
  R.Sheltami}.} \bibinfo{year}{2012}\natexlab{}.
\newblock \showarticletitle{{Context-Aware Driver Assistance System}}.
\newblock \bibinfo{journal}{\emph{Procedia Computer Science}}
  \bibinfo{volume}{10} (\bibinfo{date}{1} \bibinfo{year}{2012}),
  \bibinfo{pages}{785--794}.
\newblock
\showISSN{1877-0509}
\urldef\tempurl%
\url{https://doi.org/10.1016/J.PROCS.2012.06.100}
\showDOI{\tempurl}


\bibitem[\protect\citeauthoryear{Ali, Saifuzzaman, Zheng, and Haque}{Ali
  et~al\mbox{.}}{2018}]%
        {Ali2018}
\bibfield{author}{\bibinfo{person}{Yasir Ali}, \bibinfo{person}{Mohammad
  Saifuzzaman}, \bibinfo{person}{Zuduo Zheng}, {and}
  \bibinfo{person}{Mohammad~Mazharul Haque}.} \bibinfo{year}{2018}\natexlab{}.
\newblock \showarticletitle{{Human Factors in Modelling Mixed Traffic of
  Traditional, Connected, and Automated Vehicles}}.
\newblock In \bibinfo{booktitle}{\emph{Advances in Human Factors in Simulation
  and Modeling}}. Vol.~\bibinfo{volume}{591}. \bibinfo{pages}{262--273}.
\newblock
\showISBNx{978-3-319-60590-6}
\urldef\tempurl%
\url{https://doi.org/10.1007/978-3-319-60591-3}
\showDOI{\tempurl}


\bibitem[\protect\citeauthoryear{Antrobus, Burnett, and Krehl}{Antrobus
  et~al\mbox{.}}{2017}]%
        {Antrobus2017Driver-PassengerSystems}
\bibfield{author}{\bibinfo{person}{Vicki Antrobus}, \bibinfo{person}{Gary
  Burnett}, {and} \bibinfo{person}{Claudia Krehl}.}
  \bibinfo{year}{2017}\natexlab{}.
\newblock \showarticletitle{{Driver-Passenger Collaboration as a basis for
  Human-Machine Interface Design for Vehicle Navigation Systems}}.
\newblock \bibinfo{journal}{\emph{Ergonomics}} \bibinfo{volume}{60},
  \bibinfo{number}{3} (\bibinfo{date}{3} \bibinfo{year}{2017}),
  \bibinfo{pages}{321--332}.
\newblock
\showISSN{0014-0139}
\urldef\tempurl%
\url{https://doi.org/10.1080/00140139.2016.1172736}
\showDOI{\tempurl}


\bibitem[\protect\citeauthoryear{Attard, Haklay, and Capineri}{Attard
  et~al\mbox{.}}{2016}]%
        {Attard2016TheSystems}
\bibfield{author}{\bibinfo{person}{Maria Attard}, \bibinfo{person}{Muki
  Haklay}, {and} \bibinfo{person}{Cristina Capineri}.}
  \bibinfo{year}{2016}\natexlab{}.
\newblock \showarticletitle{{The Potential of Volunteered Geographic
  Information (VGI) in Future Transport Systems}}.
\newblock \bibinfo{journal}{\emph{Urban Planning}} \bibinfo{volume}{1},
  \bibinfo{number}{4} (\bibinfo{year}{2016}), \bibinfo{pages}{6}.
\newblock
\showISSN{2183-7635}
\urldef\tempurl%
\url{https://doi.org/10.17645/up.v1i4.612}
\showDOI{\tempurl}


\bibitem[\protect\citeauthoryear{Ben-Elia and Avineri}{Ben-Elia and
  Avineri}{2015}]%
        {Ben-Elia2015ResponseReview}
\bibfield{author}{\bibinfo{person}{Eran Ben-Elia} {and} \bibinfo{person}{Erel
  Avineri}.} \bibinfo{year}{2015}\natexlab{}.
\newblock \showarticletitle{{Response to Travel Information: A Behavioural
  Review}}.
\newblock \bibinfo{journal}{\emph{Transport Reviews}} \bibinfo{volume}{35},
  \bibinfo{number}{3} (\bibinfo{year}{2015}), \bibinfo{pages}{352--377}.
\newblock
\showISBNx{0144-1647}
\showISSN{14645327}
\urldef\tempurl%
\url{https://doi.org/10.1080/01441647.2015.1015471}
\showDOI{\tempurl}


\bibitem[\protect\citeauthoryear{Braess, Nagurney, and Wakolbinger}{Braess
  et~al\mbox{.}}{2005}]%
        {Braess2005}
\bibfield{author}{\bibinfo{person}{Dietrich Braess}, \bibinfo{person}{Anna
  Nagurney}, {and} \bibinfo{person}{Tina Wakolbinger}.}
  \bibinfo{year}{2005}\natexlab{}.
\newblock \showarticletitle{{On a Paradox of Traffic Planning}}.
\newblock \bibinfo{journal}{\emph{TRANSPORTATION SCIENCE
  Unternehmensforschung}} \bibinfo{volume}{39}, \bibinfo{number}{12}
  (\bibinfo{year}{2005}), \bibinfo{pages}{446--450}.
\newblock
\urldef\tempurl%
\url{https://doi.org/10.1287/trsc.1050.0127}
\showDOI{\tempurl}


\bibitem[\protect\citeauthoryear{Brown and Laurier}{Brown and Laurier}{2012}]%
        {Brown2012TheGPS}
\bibfield{author}{\bibinfo{person}{Barry Brown} {and} \bibinfo{person}{Eric
  Laurier}.} \bibinfo{year}{2012}\natexlab{}.
\newblock \showarticletitle{{The normal natural troubles of driving with GPS}}.
  In \bibinfo{booktitle}{\emph{Proceedings of the 2012 CHI Conference on Human
  Factors in Computing Systems - CHI '12}}. \bibinfo{publisher}{ACM},
  \bibinfo{address}{New York, New York, USA}, \bibinfo{pages}{1621--1630}.
\newblock
\showISBNx{9781450310154}
\urldef\tempurl%
\url{https://doi.org/10.1145/2207676.2208285}
\showDOI{\tempurl}


\bibitem[\protect\citeauthoryear{Chorus, Molin, and van Wee}{Chorus
  et~al\mbox{.}}{2006}]%
        {Chorus2006TravelReview}
\bibfield{author}{\bibinfo{person}{Caspar~G Chorus}, \bibinfo{person}{Eric J~E
  Molin}, {and} \bibinfo{person}{Bert van Wee}.}
  \bibinfo{year}{2006}\natexlab{}.
\newblock \showarticletitle{{Travel information as an instrument to change car
  drivers travel choices : a literature review}}.
\newblock \bibinfo{journal}{\emph{European Journal of Transport and
  Infrastructure Research}} \bibinfo{volume}{6}, \bibinfo{number}{4}
  (\bibinfo{year}{2006}), \bibinfo{pages}{335--364}.
\newblock
\showISSN{1567-7133}


\bibitem[\protect\citeauthoryear{Correll, Moritz, and Heer}{Correll
  et~al\mbox{.}}{2018}]%
        {Correll2018Value-SuppressingPalettes}
\bibfield{author}{\bibinfo{person}{Michael Correll}, \bibinfo{person}{Dominik
  Moritz}, {and} \bibinfo{person}{Jeffrey Heer}.}
  \bibinfo{year}{2018}\natexlab{}.
\newblock \showarticletitle{{Value-Suppressing Uncertainty Palettes}}. In
  \bibinfo{booktitle}{\emph{Proceedings of the 2018 CHI Conference on Human
  Factors in Computing Systems - CHI '18}}. \bibinfo{publisher}{ACM Press},
  \bibinfo{address}{New York, New York, USA}, \bibinfo{pages}{1--11}.
\newblock
\showISBNx{9781450356206}
\urldef\tempurl%
\url{https://doi.org/10.1145/3173574.3174216}
\showDOI{\tempurl}


\bibitem[\protect\citeauthoryear{Dingus, Hulse, Mollenhauer, Fleischman,
  Mcgehee, and Manakkal}{Dingus et~al\mbox{.}}{1997}]%
        {Dingus1997a}
\bibfield{author}{\bibinfo{person}{Thomas~A. Dingus},
  \bibinfo{person}{Melissa~C. Hulse}, \bibinfo{person}{Michael~A. Mollenhauer},
  \bibinfo{person}{Rebecca~N. Fleischman}, \bibinfo{person}{Daniel~V. Mcgehee},
  {and} \bibinfo{person}{Natarajan Manakkal}.} \bibinfo{year}{1997}\natexlab{}.
\newblock \showarticletitle{{Effects of Age, System Experience, and Navigation
  Technique on Driving with an Advanced Traveler Information System}}.
\newblock \bibinfo{journal}{\emph{Human Factors: The Journal of the Human
  Factors and Ergonomics Society}} \bibinfo{volume}{39}, \bibinfo{number}{2}
  (\bibinfo{year}{1997}), \bibinfo{pages}{177--199}.
\newblock
\showISBNx{00187208 (ISSN)}
\showISSN{0018-7208}
\urldef\tempurl%
\url{https://doi.org/10.1518/001872097778543804}
\showDOI{\tempurl}


\bibitem[\protect\citeauthoryear{Ekstrand, Harper, Willemsen, and
  Konstan}{Ekstrand et~al\mbox{.}}{2014}]%
        {Ekstrand2014UserAlgorithms}
\bibfield{author}{\bibinfo{person}{Michael~D. Ekstrand},
  \bibinfo{person}{F.~Maxwell Harper}, \bibinfo{person}{Martijn~C. Willemsen},
  {and} \bibinfo{person}{Joseph~A. Konstan}.} \bibinfo{year}{2014}\natexlab{}.
\newblock \showarticletitle{{User Perception of Differences in Recommender
  Algorithms}}. In \bibinfo{booktitle}{\emph{Proceedings of the 8th ACM
  Conference on Recommender systems - RecSys '14}}. \bibinfo{publisher}{ACM
  Press}, \bibinfo{address}{New York, New York, USA},
  \bibinfo{pages}{161--168}.
\newblock
\showISBNx{9781450326681}
\urldef\tempurl%
\url{https://doi.org/10.1145/2645710.2645737}
\showDOI{\tempurl}


\bibitem[\protect\citeauthoryear{Fernandes, Walls, Munson, Hullman, and
  Kay}{Fernandes et~al\mbox{.}}{2018}]%
        {Fernandes2018UncertaintyDecision-Making}
\bibfield{author}{\bibinfo{person}{Michael Fernandes}, \bibinfo{person}{Logan
  Walls}, \bibinfo{person}{Sean Munson}, \bibinfo{person}{Jessica Hullman},
  {and} \bibinfo{person}{Matthew Kay}.} \bibinfo{year}{2018}\natexlab{}.
\newblock \showarticletitle{{Uncertainty Displays Using Quantile Dotplots or
  CDFs Improve Transit Decision-Making}}. In
  \bibinfo{booktitle}{\emph{Proceedings of the 2018 CHI Conference on Human
  Factors in Computing Systems - CHI '18}}. \bibinfo{publisher}{ACM Press},
  \bibinfo{address}{New York, New York, USA}, \bibinfo{pages}{1--12}.
\newblock
\showISBNx{9781450356206}
\urldef\tempurl%
\url{https://doi.org/10.1145/3173574.3173718}
\showDOI{\tempurl}


\bibitem[\protect\citeauthoryear{Forlizzi, Barley, and Seder}{Forlizzi
  et~al\mbox{.}}{2010}]%
        {Forlizzi2010WhereTurn}
\bibfield{author}{\bibinfo{person}{Jodi Forlizzi}, \bibinfo{person}{William~C.
  Barley}, {and} \bibinfo{person}{Thomas Seder}.}
  \bibinfo{year}{2010}\natexlab{}.
\newblock \showarticletitle{{Where should i turn}}. In
  \bibinfo{booktitle}{\emph{Proceedings of the 28th international conference on
  Human factors in computing systems - CHI '10}}. \bibinfo{publisher}{ACM
  Press}, \bibinfo{address}{New York, New York, USA}, \bibinfo{pages}{1261}.
\newblock
\showISBNx{9781605589299}
\urldef\tempurl%
\url{https://doi.org/10.1145/1753326.1753516}
\showDOI{\tempurl}


\bibitem[\protect\citeauthoryear{Fujino, Hashimoto, Kasahara, Mori, Iiyama, and
  Minoh}{Fujino et~al\mbox{.}}{2018}]%
        {Fujino2018DetectingTracks}
\bibfield{author}{\bibinfo{person}{Takumi Fujino}, \bibinfo{person}{Atsushi
  Hashimoto}, \bibinfo{person}{Hidekazu Kasahara}, \bibinfo{person}{Mikihiko
  Mori}, \bibinfo{person}{Masaaki Iiyama}, {and} \bibinfo{person}{Michihiko
  Minoh}.} \bibinfo{year}{2018}\natexlab{}.
\newblock \showarticletitle{{Detecting Deviations from Intended Routes Using
  Vehicular GPS Tracks}}.
\newblock \bibinfo{journal}{\emph{ACM Transactions on Spatial Algorithms and
  Systems}} \bibinfo{volume}{4}, \bibinfo{number}{1} (\bibinfo{date}{6}
  \bibinfo{year}{2018}), \bibinfo{pages}{1--21}.
\newblock
\showISSN{23740353}
\urldef\tempurl%
\url{https://doi.org/10.1145/3204455}
\showDOI{\tempurl}


\bibitem[\protect\citeauthoryear{Hu and Pu}{Hu and Pu}{2010}]%
        {Hu2010ASystems}
\bibfield{author}{\bibinfo{person}{Rong Hu} {and} \bibinfo{person}{Pearl Pu}.}
  \bibinfo{year}{2010}\natexlab{}.
\newblock \showarticletitle{{A Study on User Perception of Personality-Based
  Recommender Systems}}.
\newblock In \bibinfo{booktitle}{\emph{User Modeling, Adaptation, and
  Personalization. UMAP 2010. Lecture Notes in Computer Science, vol 6075}}.
  \bibinfo{pages}{291--302}.
\newblock
\urldef\tempurl%
\url{https://doi.org/10.1007/978-3-642-13470-8{\_}27}
\showDOI{\tempurl}


\bibitem[\protect\citeauthoryear{{J.D. Power}}{{J.D. Power}}{2012}]%
        {J.D.Power2012VehicleDeclines}
\bibfield{author}{\bibinfo{person}{{J.D. Power}}.}
  \bibinfo{year}{2012}\natexlab{}.
\newblock \bibinfo{booktitle}{\emph{{Vehicle Owners Ask for Smartphone
  Integration and Better Voice Controls, as Satisfaction with Factory-Installed
  Navigation Systems Declines}}}.
\newblock \bibinfo{type}{{T}echnical {R}eport}. \bibinfo{institution}{J.D.
  Power}.
\newblock
\urldef\tempurl%
\url{www.jdpower.com/corporate}
\showURL{%
\tempurl}


\bibitem[\protect\citeauthoryear{{J.D. Power}}{{J.D. Power}}{2017}]%
        {J.D.Power2017ImprovementsFinds}
\bibfield{author}{\bibinfo{person}{{J.D. Power}}.}
  \bibinfo{year}{2017}\natexlab{}.
\newblock \bibinfo{booktitle}{\emph{{Improvements Needed on Navigation Systems,
  J.D. Power Finds}}}.
\newblock \bibinfo{type}{{T}echnical {R}eport}. \bibinfo{institution}{J.D.
  Power}.
\newblock
\urldef\tempurl%
\url{www.jdpower.com/about-us/press-release-info}
\showURL{%
\tempurl}


\bibitem[\protect\citeauthoryear{Karatas, Yoshikawa, De~Silva, and
  Okada}{Karatas et~al\mbox{.}}{2016}]%
        {Karatas2016NAMIDA:Driver}
\bibfield{author}{\bibinfo{person}{Nihan Karatas}, \bibinfo{person}{Soshi
  Yoshikawa}, \bibinfo{person}{P.~Ravindra De~Silva}, {and}
  \bibinfo{person}{Michio Okada}.} \bibinfo{year}{2016}\natexlab{}.
\newblock \showarticletitle{{NAMIDA: How to Reduce the Cognitive Workload of
  Driver}}. In \bibinfo{booktitle}{\emph{The Eleventh ACM/IEEE International
  Conference on Human Robot Interaction}}. \bibinfo{publisher}{IEEE Press},
  \bibinfo{pages}{651}.
\newblock
\showISBNx{9781467383707}
\urldef\tempurl%
\url{https://dl.acm.org/citation.cfm?id=2906921}
\showURL{%
\tempurl}


\bibitem[\protect\citeauthoryear{Knijnenburg, Willemsen, Gantner, Soncu, and
  Newell}{Knijnenburg et~al\mbox{.}}{2012}]%
        {Knijnenburg2012ExplainingSystems}
\bibfield{author}{\bibinfo{person}{Bart~P Knijnenburg},
  \bibinfo{person}{Martijn~C Willemsen}, \bibinfo{person}{Zeno Gantner},
  \bibinfo{person}{Hakan Soncu}, {and} \bibinfo{person}{Chris Newell}.}
  \bibinfo{year}{2012}\natexlab{}.
\newblock \showarticletitle{{Explaining the user experience of recommender
  systems}}.
\newblock \bibinfo{journal}{\emph{User Modeling and User-Adapted Interaction}}
  \bibinfo{volume}{22} (\bibinfo{year}{2012}), \bibinfo{pages}{441--504}.
\newblock
\urldef\tempurl%
\url{https://doi.org/10.1007/s11257-011-9118-4}
\showDOI{\tempurl}


\bibitem[\protect\citeauthoryear{Levine, Shinar, and Shabtai}{Levine
  et~al\mbox{.}}{2014}]%
        {Levine2014SystemExchange}
\bibfield{author}{\bibinfo{person}{Uri Levine}, \bibinfo{person}{Amir Shinar},
  {and} \bibinfo{person}{Ehud Shabtai}.} \bibinfo{year}{2014}\natexlab{}.
\newblock \bibinfo{title}{{System and Method for Realtime Community Information
  Exchange}}.
\newblock , \bibinfo{numpages}{12}~pages.
\newblock
\urldef\tempurl%
\url{https://patentimages.storage.googleapis.com/8d/00/0a/75de486f22fbc1/US8762035.pdf}
\showURL{%
\tempurl}


\bibitem[\protect\citeauthoryear{Lin, Hsu, Talamonti, Zhang, Oney, Mars, and
  Tang}{Lin et~al\mbox{.}}{2018}]%
        {Lin2018Adasa}
\bibfield{author}{\bibinfo{person}{Shih-Chieh Lin}, \bibinfo{person}{Chang-Hong
  Hsu}, \bibinfo{person}{Walter Talamonti}, \bibinfo{person}{Yunqi Zhang},
  \bibinfo{person}{Steve Oney}, \bibinfo{person}{Jason Mars}, {and}
  \bibinfo{person}{Lingjia Tang}.} \bibinfo{year}{2018}\natexlab{}.
\newblock \showarticletitle{{Adasa}}. In \bibinfo{booktitle}{\emph{The 31st
  Annual ACM Symposium on User Interface Software and Technology - UIST '18}}.
  \bibinfo{publisher}{ACM Press}, \bibinfo{address}{New York, New York, USA},
  \bibinfo{pages}{531--542}.
\newblock
\showISBNx{9781450359481}
\urldef\tempurl%
\url{https://doi.org/10.1145/3242587.3242593}
\showDOI{\tempurl}


\bibitem[\protect\citeauthoryear{Mahmud, Mubin, and Shahid}{Mahmud
  et~al\mbox{.}}{2009}]%
        {Mahmud2009UserDrivers}
\bibfield{author}{\bibinfo{person}{Abdullah~Al Mahmud}, \bibinfo{person}{Omar
  Mubin}, {and} \bibinfo{person}{Suleman Shahid}.}
  \bibinfo{year}{2009}\natexlab{}.
\newblock \showarticletitle{{User experience with in-car GPS navigation
  systems: comparing the young and elderly drivers}}. In
  \bibinfo{booktitle}{\emph{Proceedings of the 11th International Conference on
  Human-Computer Interaction with Mobile Devices and Services}}.
  \bibinfo{publisher}{ACM}, \bibinfo{pages}{1--90}.
\newblock
\showISBNx{9781605582818}
\urldef\tempurl%
\url{https://doi.org/10.1145/1613858.1613962}
\showDOI{\tempurl}


\bibitem[\protect\citeauthoryear{Mehndiratta and Quiros}{Mehndiratta and
  Quiros}{2017}]%
        {Mehndiratta2017}
\bibfield{author}{\bibinfo{person}{Shomik Mehndiratta} {and}
  \bibinfo{person}{Tatiana~Peralta Quiros}.} \bibinfo{year}{2017}\natexlab{}.
\newblock \bibinfo{title}{{Traffic jams, pollution, road crashes: Can
  technology end the woes of urban transport?}}
\newblock
\newblock
\urldef\tempurl%
\url{http://blogs.worldbank.org/transport/traffic-jams-pollution-road-crashes-can-technology-end-woes-urban-transport}
\showURL{%
\tempurl}


\bibitem[\protect\citeauthoryear{Mikami}{Mikami}{1978}]%
        {Mikami1978CACS-UrbanControl}
\bibfield{author}{\bibinfo{person}{Toru Mikami}.}
  \bibinfo{year}{1978}\natexlab{}.
\newblock \showarticletitle{{CACS-Urban traffic control system featuring
  computer control}}. In \bibinfo{booktitle}{\emph{National Computer
  Conference}}.
\newblock
\urldef\tempurl%
\url{www.computerhistory.org}
\showURL{%
\tempurl}


\bibitem[\protect\citeauthoryear{Monreal and Rossettit}{Monreal and
  Rossettit}{2014}]%
        {Monreal2014}
\bibfield{author}{\bibinfo{person}{Cristina~Olaverri Monreal} {and}
  \bibinfo{person}{Rosaldo J.~F. Rossettit}.} \bibinfo{year}{2014}\natexlab{}.
\newblock \showarticletitle{{Human Factors in Intelligent Transportation
  Systems}}.
\newblock \bibinfo{journal}{\emph{IEEE TRANSACTIONS ON INTELLIGENT
  TRANSPORTATION SYSTEMS}} \bibinfo{volume}{15}, \bibinfo{number}{4}
  (\bibinfo{year}{2014}), \bibinfo{pages}{480}.
\newblock
\showISBNx{1317781104}


\bibitem[\protect\citeauthoryear{Muller}{Muller}{2014}]%
        {Muller2014CuriosityMethod}
\bibfield{author}{\bibinfo{person}{Michael Muller}.}
  \bibinfo{year}{2014}\natexlab{}.
\newblock \showarticletitle{{Curiosity, Creativity, and Surprise as Analytic
  Tools: Grounded Theory Method}}.
\newblock In \bibinfo{booktitle}{\emph{Ways of Knowing in HCI}}.
  \bibinfo{publisher}{Springer New York}, \bibinfo{address}{New York, NY},
  \bibinfo{pages}{25--48}.
\newblock
\urldef\tempurl%
\url{https://doi.org/10.1007/978-1-4939-0378-8{\_}2}
\showDOI{\tempurl}


\bibitem[\protect\citeauthoryear{Muller, Guha, Baumer, Mimno, and Shami}{Muller
  et~al\mbox{.}}{2016}]%
        {Muller2016MachineMethod}
\bibfield{author}{\bibinfo{person}{Michael Muller}, \bibinfo{person}{Shion
  Guha}, \bibinfo{person}{Eric~P.S. Baumer}, \bibinfo{person}{David Mimno},
  {and} \bibinfo{person}{N.~Sadat Shami}.} \bibinfo{year}{2016}\natexlab{}.
\newblock \showarticletitle{{Machine Learning and Grounded Theory Method}}. In
  \bibinfo{booktitle}{\emph{Proceedings of the 19th International Conference on
  Supporting Group Work - GROUP '16}}. \bibinfo{publisher}{ACM Press},
  \bibinfo{address}{New York, New York, USA}, \bibinfo{pages}{3--8}.
\newblock
\showISBNx{9781450342766}
\urldef\tempurl%
\url{https://doi.org/10.1145/2957276.2957280}
\showDOI{\tempurl}


\bibitem[\protect\citeauthoryear{Patel, Chen, Smith, and Landay}{Patel
  et~al\mbox{.}}{2006}]%
        {Patel2006PersonalizingRoutes}
\bibfield{author}{\bibinfo{person}{Kayur Patel}, \bibinfo{person}{Mike~Y.
  Chen}, \bibinfo{person}{Ian Smith}, {and} \bibinfo{person}{James~A. Landay}.}
  \bibinfo{year}{2006}\natexlab{}.
\newblock \showarticletitle{{Personalizing routes}}. In
  \bibinfo{booktitle}{\emph{Proceedings of the 19th annual ACM symposium on
  User interface software and technology - UIST '06}}.
  \bibinfo{publisher}{ACM}, \bibinfo{address}{New York, New York, USA},
  \bibinfo{pages}{187--190}.
\newblock
\showISBNx{1595933131}
\urldef\tempurl%
\url{https://doi.org/10.1145/1166253.1166282}
\showDOI{\tempurl}


\bibitem[\protect\citeauthoryear{Peterson}{Peterson}{2018}]%
        {Peterson2018}
\bibfield{author}{\bibinfo{person}{David Peterson}.}
  \bibinfo{year}{2018}\natexlab{}.
\newblock \bibinfo{title}{{Can you overcome the Einstellung Effect? - David L.
  Peterson -}}.
\newblock
\newblock
\urldef\tempurl%
\url{https://davidpeterson.com/can-overcome-einstellung-effect/}
\showURL{%
\tempurl}


\bibitem[\protect\citeauthoryear{Pfleging, Meschtscherjakov, Schneegass, and
  Tscheligi}{Pfleging et~al\mbox{.}}{2014}]%
        {Pfleging2014ExperienceNavigation}
\bibfield{author}{\bibinfo{person}{Bastian Pfleging},
  \bibinfo{person}{Alexander Meschtscherjakov}, \bibinfo{person}{Stefan
  Schneegass}, {and} \bibinfo{person}{Manfred Tscheligi}.}
  \bibinfo{year}{2014}\natexlab{}.
\newblock \showarticletitle{{Experience Maps: Experience-Enhanced Routes for
  Car Navigation}}. In \bibinfo{booktitle}{\emph{Proceedings of the 6th
  International Conference on Automotive User Interfaces and Interactive
  Vehicular Applications - AutomotiveUI '14}}. \bibinfo{publisher}{ACM},
  \bibinfo{address}{New York, New York, USA}, \bibinfo{pages}{1--6}.
\newblock
\showISBNx{9781450307253}
\urldef\tempurl%
\url{https://doi.org/10.1145/2667239.2667275}
\showDOI{\tempurl}


\bibitem[\protect\citeauthoryear{{Qing Yang} and {Honggang Wang}}{{Qing Yang}
  and {Honggang Wang}}{2015}]%
        {QingYang2015TowardNetworks}
\bibfield{author}{\bibinfo{person}{{Qing Yang}} {and}
  \bibinfo{person}{{Honggang Wang}}.} \bibinfo{year}{2015}\natexlab{}.
\newblock \showarticletitle{{Toward trustworthy vehicular social networks}}.
\newblock \bibinfo{journal}{\emph{IEEE Communications Magazine}}
  \bibinfo{volume}{53}, \bibinfo{number}{8} (\bibinfo{date}{8}
  \bibinfo{year}{2015}), \bibinfo{pages}{42--47}.
\newblock
\showISSN{0163-6804}
\urldef\tempurl%
\url{https://doi.org/10.1109/MCOM.2015.7180506}
\showDOI{\tempurl}


\bibitem[\protect\citeauthoryear{Quercia, Schifanella, and Aiello}{Quercia
  et~al\mbox{.}}{2014}]%
        {Quercia2014}
\bibfield{author}{\bibinfo{person}{Daniele Quercia}, \bibinfo{person}{Rossano
  Schifanella}, {and} \bibinfo{person}{Luca~Maria Aiello}.}
  \bibinfo{year}{2014}\natexlab{}.
\newblock \showarticletitle{{The shortest path to happiness}}. In
  \bibinfo{booktitle}{\emph{Proceedings of the 25th ACM conference on Hypertext
  and social media - HT '14}}. \bibinfo{publisher}{ACM Press},
  \bibinfo{address}{New York, New York, USA}, \bibinfo{pages}{116--125}.
\newblock
\showISBNx{9781450329545}
\urldef\tempurl%
\url{https://doi.org/10.1145/2631775.2631799}
\showDOI{\tempurl}


\bibitem[\protect\citeauthoryear{Sakamoto, Komatsu, and Igarashi}{Sakamoto
  et~al\mbox{.}}{2013}]%
        {Sakamoto2013VoiceManipulation}
\bibfield{author}{\bibinfo{person}{Daisuke Sakamoto}, \bibinfo{person}{Takanori
  Komatsu}, {and} \bibinfo{person}{Takeo Igarashi}.}
  \bibinfo{year}{2013}\natexlab{}.
\newblock \showarticletitle{{Voice augmented manipulation}}. In
  \bibinfo{booktitle}{\emph{Proceedings of the 15th international conference on
  Human-computer interaction with mobile devices and services - MobileHCI
  '13}}. \bibinfo{publisher}{ACM Press}, \bibinfo{address}{New York, New York,
  USA}, \bibinfo{pages}{69}.
\newblock
\showISBNx{9781450322737}
\urldef\tempurl%
\url{https://doi.org/10.1145/2493190.2493244}
\showDOI{\tempurl}


\bibitem[\protect\citeauthoryear{Sha, Kwak, Nath, and Iftode}{Sha
  et~al\mbox{.}}{2013}]%
        {Sha2013SocialNavigation}
\bibfield{author}{\bibinfo{person}{Wenjie Sha}, \bibinfo{person}{Daehan Kwak},
  \bibinfo{person}{Badri Nath}, {and} \bibinfo{person}{Liviu Iftode}.}
  \bibinfo{year}{2013}\natexlab{}.
\newblock \showarticletitle{{Social vehicle navigation: integrating shared
  driving experience into vehicle navigation}}. In
  \bibinfo{booktitle}{\emph{Proceedings of the 14th Workshop on Mobile
  Computing Systems and Applications - HotMobile '13}}. \bibinfo{publisher}{ACM
  Press}, \bibinfo{address}{New York, New York, USA}, \bibinfo{pages}{1}.
\newblock
\showISBNx{9781450314213}
\urldef\tempurl%
\url{https://doi.org/10.1145/2444776.2444798}
\showDOI{\tempurl}


\bibitem[\protect\citeauthoryear{Silva, Celes, Neto, Mota, da~Cunha, Ferreira,
  Ribeiro, Vaz~de Melo, Almeida, and Loureiro}{Silva et~al\mbox{.}}{2016}]%
        {Silva2016UsersOpportunities}
\bibfield{author}{\bibinfo{person}{T.H. Silva}, \bibinfo{person}{C.S.F.S.
  Celes}, \bibinfo{person}{J.B.B. Neto}, \bibinfo{person}{V.F.S. Mota},
  \bibinfo{person}{F.D. da Cunha}, \bibinfo{person}{A.P.G. Ferreira},
  \bibinfo{person}{A.I.J.T. Ribeiro}, \bibinfo{person}{P.O.S. Vaz~de Melo},
  \bibinfo{person}{J.M. Almeida}, {and} \bibinfo{person}{A.A.F. Loureiro}.}
  \bibinfo{year}{2016}\natexlab{}.
\newblock \bibinfo{booktitle}{\emph{{Users in the urban sensing process:
  Challenges and research opportunities}}}.
\newblock \bibinfo{publisher}{Elsevier Inc.} 45--95 pages.
\newblock
\showISBNx{9780128037027}
\urldef\tempurl%
\url{https://doi.org/10.1016/B978-0-12-803663-1.00003-6}
\showDOI{\tempurl}


\bibitem[\protect\citeauthoryear{Silva, Vaz De~Melo, Viana, Almeida, Salles,
  and Loureiro}{Silva et~al\mbox{.}}{2013}]%
        {Silva2013TrafficAlerts}
\bibfield{author}{\bibinfo{person}{Thiago~H. Silva},
  \bibinfo{person}{Pedro~O.S. Vaz De~Melo}, \bibinfo{person}{Aline~Carneiro
  Viana}, \bibinfo{person}{Jussara~M. Almeida}, \bibinfo{person}{Juliana
  Salles}, {and} \bibinfo{person}{Antonio~A.F. Loureiro}.}
  \bibinfo{year}{2013}\natexlab{}.
\newblock \showarticletitle{{Traffic condition is more than colored lines on a
  map: Characterization of Waze alerts}}.
\newblock \bibinfo{journal}{\emph{Lecture Notes in Computer Science (including
  subseries Lecture Notes in Artificial Intelligence and Lecture Notes in
  Bioinformatics)}}  \bibinfo{volume}{8238 LNCS} (\bibinfo{year}{2013}),
  \bibinfo{pages}{309--318}.
\newblock
\showISBNx{9783319032597}
\showISSN{03029743}
\urldef\tempurl%
\url{https://doi.org/10.1007/978-3-319-03260-3{\_}27}
\showDOI{\tempurl}


\bibitem[\protect\citeauthoryear{Soegaard and Dam}{Soegaard and Dam}{2013}]%
        {Soegaard2013}
\bibfield{editor}{\bibinfo{person}{Mads Soegaard} {and}
  \bibinfo{person}{Rikke~Friis Dam}} (Eds.). \bibinfo{year}{2013}\natexlab{}.
\newblock \bibinfo{booktitle}{\emph{{The Encyclopedia of Human-Computer
  Interaction}} (\bibinfo{edition}{2} ed.)}.
\newblock \bibinfo{publisher}{Interaction Design Foundation}.
\newblock
\showISBNx{9788792964}
\urldef\tempurl%
\url{https://www.interaction-design.org/literature/book/the-encyclopedia-of-human-computer-interaction-2nd-ed}
\showURL{%
\tempurl}


\bibitem[\protect\citeauthoryear{Tang and Cheng}{Tang and Cheng}{2016}]%
        {Tang2016AnalyzingData}
\bibfield{author}{\bibinfo{person}{W. Tang} {and} \bibinfo{person}{L. Cheng}.}
  \bibinfo{year}{2016}\natexlab{}.
\newblock \showarticletitle{{Analyzing multiday route choice behavior of
  commuters using GPS data}}.
\newblock \bibinfo{journal}{\emph{Advances in Mechanical Engineering}}
  \bibinfo{volume}{8}, \bibinfo{number}{2} (\bibinfo{year}{2016}),
  \bibinfo{pages}{1--11}.
\newblock
\showISBNx{1687814016633}
\showISSN{1687-8140}
\urldef\tempurl%
\url{https://doi.org/10.1177/1687814016633030}
\showDOI{\tempurl}


\bibitem[\protect\citeauthoryear{{United Nations}}{{United Nations}}{2017}]%
        {UnitedNations2017}
\bibfield{author}{\bibinfo{person}{{United Nations}}.}
  \bibinfo{year}{2017}\natexlab{}.
\newblock \showarticletitle{{Progress towards the Sustainable Development
  Goals}}.
\newblock \bibinfo{journal}{\emph{Report of the Secretary-General}}
  \bibinfo{volume}{E/2017/66}, \bibinfo{number}{May} (\bibinfo{year}{2017}),
  \bibinfo{pages}{19}.
\newblock
\showISBNx{GE.02-10152 (E) 040302}
\showISSN{0020-8183}
\urldef\tempurl%
\url{https://doi.org/10.1017/S0020818300006640}
\showDOI{\tempurl}


\bibitem[\protect\citeauthoryear{Valdes-Dapena}{Valdes-Dapena}{2016}]%
        {Valdes-Dapena2016MostDirections}
\bibfield{author}{\bibinfo{person}{Peter Valdes-Dapena}.}
  \bibinfo{year}{2016}\natexlab{}.
\newblock \bibinfo{title}{{Most drivers who own cars with built-in GPS systems
  use phones for directions}}.
\newblock
\newblock
\urldef\tempurl%
\url{http://money.cnn.com/2016/10/10/autos/car-navigation-frustration/index.html}
\showURL{%
\tempurl}


\bibitem[\protect\citeauthoryear{Vyroubal, Stancic, and Grgurevic}{Vyroubal
  et~al\mbox{.}}{2016}]%
        {Vyroubal2016MobileSystems}
\bibfield{author}{\bibinfo{person}{Vedran Vyroubal}, \bibinfo{person}{Adam
  Stancic}, {and} \bibinfo{person}{Ivan Grgurevic}.}
  \bibinfo{year}{2016}\natexlab{}.
\newblock \showarticletitle{{Mobile devices as authentic and trustworthy
  sources in multi-agent systems}}. In \bibinfo{booktitle}{\emph{2016 39th
  International Convention on Information and Communication Technology,
  Electronics and Microelectronics (MIPRO)}}. \bibinfo{publisher}{IEEE},
  \bibinfo{pages}{661--666}.
\newblock
\showISBNx{978-953-233-086-1}
\urldef\tempurl%
\url{https://doi.org/10.1109/MIPRO.2016.7522223}
\showDOI{\tempurl}


\bibitem[\protect\citeauthoryear{Wang, Li, Sakamoto, and Igarashi}{Wang
  et~al\mbox{.}}{2014}]%
        {Wang2014HierarchicalNavigation}
\bibfield{author}{\bibinfo{person}{Fangzhou Wang}, \bibinfo{person}{Yang Li},
  \bibinfo{person}{Daisuke Sakamoto}, {and} \bibinfo{person}{Takeo Igarashi}.}
  \bibinfo{year}{2014}\natexlab{}.
\newblock \showarticletitle{{Hierarchical route maps for efficient
  navigation}}. In \bibinfo{booktitle}{\emph{Proceedings of the 19th
  international conference on Intelligent User Interfaces - IUI '14}}.
  \bibinfo{publisher}{ACM Press}, \bibinfo{address}{New York, New York, USA},
  \bibinfo{pages}{169--178}.
\newblock
\showISBNx{9781450321846}
\urldef\tempurl%
\url{https://doi.org/10.1145/2557500.2557514}
\showDOI{\tempurl}


\bibitem[\protect\citeauthoryear{{Waze}}{{Waze}}{2016}]%
        {Waze2016DriverIndex}
\bibfield{author}{\bibinfo{person}{{Waze}}.} \bibinfo{year}{2016}\natexlab{}.
\newblock \bibinfo{booktitle}{\emph{{Driver Satisfaction Index}}}.
\newblock \bibinfo{type}{{T}echnical {R}eport}. \bibinfo{institution}{Waze}.
\newblock
\urldef\tempurl%
\url{https://inbox-static.waze.com/driverindex.pdf}
\showURL{%
\tempurl}


\bibitem[\protect\citeauthoryear{Williams, Flores, and Peters}{Williams
  et~al\mbox{.}}{2014}]%
        {Williams2014AffectiveSociability}
\bibfield{author}{\bibinfo{person}{Kenton Williams},
  \bibinfo{person}{José~Acevedo Flores}, {and} \bibinfo{person}{Joshua
  Peters}.} \bibinfo{year}{2014}\natexlab{}.
\newblock \showarticletitle{{Affective Robot Influence on Driver Adherence to
  Safety, Cognitive Load Reduction and Sociability}}. In
  \bibinfo{booktitle}{\emph{Proceedings of the 6th International Conference on
  Automotive User Interfaces and Interactive Vehicular Applications -
  AutomotiveUI '14}}. \bibinfo{publisher}{ACM Press}, \bibinfo{address}{New
  York, New York, USA}, \bibinfo{pages}{1--8}.
\newblock
\showISBNx{9781450332125}
\urldef\tempurl%
\url{https://doi.org/10.1145/2667317.2667342}
\showDOI{\tempurl}


\bibitem[\protect\citeauthoryear{Wu}{Wu}{2015}]%
        {Wu2015HybridSystems}
\bibfield{author}{\bibinfo{person}{Mengqi Wu}.}
  \bibinfo{year}{2015}\natexlab{}.
\newblock \emph{\bibinfo{title}{{Hybrid user perception model: comparing users'
  perceptions toward collaborative, content-based, and hybrid recommender
  systems}}}.
\newblock \bibinfo{thesistype}{Ph.D. Dissertation}. \bibinfo{school}{Iowa State
  University}.
\newblock
\urldef\tempurl%
\url{https://lib.dr.iastate.edu/etd/14739}
\showURL{%
\tempurl}


\bibitem[\protect\citeauthoryear{Xie and Wang}{Xie and Wang}{2015}]%
        {Xie2015AnNetworks}
\bibfield{author}{\bibinfo{person}{Xiao-feng Xie} {and}
  \bibinfo{person}{Zun-jing Wang}.} \bibinfo{year}{2015}\natexlab{}.
\newblock \showarticletitle{{An Empirical Study of Combining Participatory and
  Physical Sensing to Better Understand and Improve Urban Mobility Networks}}.
\newblock \bibinfo{journal}{\emph{Transportation Research Board 94th Annual
  Meeting}} (\bibinfo{year}{2015}).
\newblock


\bibitem[\protect\citeauthoryear{Zhu and Levinson}{Zhu and Levinson}{2015}]%
        {Zhu2015DoPrinciple}
\bibfield{author}{\bibinfo{person}{Shanjiang Zhu} {and} \bibinfo{person}{David
  Levinson}.} \bibinfo{year}{2015}\natexlab{}.
\newblock \showarticletitle{{Do People Use the Shortest Path? An Empirical Test
  of Wardrop's First Principle}}.
\newblock \bibinfo{journal}{\emph{PLoS ONE}} (\bibinfo{year}{2015}).
\newblock
\urldef\tempurl%
\url{https://doi.org/10.1371/journal.pone.0134322}
\showDOI{\tempurl}


\end{thebibliography}

\end{document}